\documentclass[11pt,a4paper]{article}
\pdfoutput=1
\usepackage[left=1in,right=1in,top=1in,bottom=1in]{geometry}
\usepackage{graphicx}
\usepackage{float}
\usepackage{subcaption}
\usepackage[tbtags]{amsmath}
\usepackage{amssymb}
\usepackage{amsthm}
\usepackage{hyperref}
\usepackage[numbers]{natbib}
\usepackage{tikz}
\usepackage{datetime}\usdate
\usepackage{mdframed}
\mdfsetup{backgroundcolor=white,nobreak=true,footnoteinside=false}
\newtheorem{theorem}{Theorem}
\newtheorem{lemma}{Lemma}
\theoremstyle{definition}
\newtheorem{definition}{Definition}
\def \R{\mathbb R}
\newcommand{\sset}[1]{\left\{ #1\right\}}
\newcommand{\ssets}[1]{\{ #1\}}
\newcommand{\fwh}[1]{\; \left| \; #1 \right.}

\DeclareMathOperator*{\argmax}{argmax}
\newcommand{\algoname}[1]{\ensuremath{\text{\rm\sc #1}}}

\newcommand{\bigunion}{\bigcup}    
\newcommand{\map}{\longrightarrow}
 
\newcommand{\inters}{\cap} 
\newcommand{\then}{\Longrightarrow} 
\DeclareMathOperator*{\expect}{\mathbb E}
\newcommand{\vecc}[1]{\ensuremath{\mathbf{#1}}}

\newcommand{\revenue}[1]{\ensuremath{\text{\rm\sc Rev}\left(#1\right)}}
\title{Bounding the Optimal Revenue of Selling Multiple Goods\footnote{Supported by the European Union FP7-ICT grant 284731 (UaESMC) and ERC Advanced Grant 321171 (ALGAME).}}
\author{Yiannis Giannakopoulos\thanks{Department of Computer Science, University of Oxford. Email: \href{mailto:ygiannak@cs.ox.ac.uk}{\nolinkurl{ygiannak@cs.ox.ac.uk} }} }
\date{March 12, 2015}
\begin{document}
\maketitle
\begin{abstract}
Using duality theory techniques we derive simple, closed-form formulas for bounding the optimal revenue of a monopolist selling many heterogeneous goods, in the case where the buyer's valuations for the items come i.i.d.\ from a uniform distribution and in the case where they follow independent (but not necessarily identical) exponential distributions. We apply this in order to get in both these settings specific performance guarantees, as functions of the number of items $m$, for the simple deterministic selling mechanisms studied by~\citet{Hart:2012uq}, namely the one that sells the items separately and the one that offers them all in a single bundle.

We also propose and study the performance of a natural randomized mechanism for exponential valuations, called \algoname{Proportional}. As an interesting corollary, for the special case where the exponential distributions are also identical, we can derive that offering the goods in a single full bundle is the optimal selling mechanism for \emph{any number of items}. To our knowledge, this is the first result of its kind: finding a revenue-maximizing auction in an additive setting with arbitrarily many goods.
\end{abstract}

\section{Introduction}
\label{sec:intro}
What selling mechanism should a monopolist with many heterogeneous goods deploy when facing a buyer, in order to maximize his revenue?
The buyer has private valuations for the items but the seller can only have an incomplete prior knowledge of these values, in the form of a probability distribution over them. Furthermore, the buyer is strategic and selfish, meaning that if asked to submit her valuations for the goods she will lie if this is to improve her own personal gain. This is one of the most fundamental problems in auction theory, and although it has received a lot of attention from both the Economics and Computer Science communities, it still remains elusive.

\subsection{Related Work}
\label{sec:related-work}
The problem is famous for demonstrating a striking dichotomy; the case of a single good is fully resolved by the seminal work of~\citet{Myerson:1981aa}: the optimal selling strategy is to just set a take-it-or-leave it price for the item, and this price is given by a very simple formula involving the probability distribution. However, for more goods, in fact even for just two and three items, our understanding of optimal mechanisms is totally insufficient. The problem seems to be qualitatively completely different and it is widely believed that no closed-form, elegant solutions are within reach (see e.g.~\cite{Manelli:2007kx,Daskalakis:2012fk}).

First of all, it is known that in general determinism (i.e.\ setting selling prices for various bundles of goods) is not enough any more and that lotteries need to be deployed for optimality~\citep{Hart:2012zr,Hart:2012ys,Manelli:2006vn,Pycia:2006pd,Daskalakis:2013vn}.
\citet{Manelli:2006vn} provide some sufficient conditions for deterministic mechanisms to be optimal, but these are arguably rather involved, so they were able to instantiate and verify them only for the case of two and three goods with valuations i.i.d.\ according to a uniform distribution. \citet{Hart:2012uq} have also presented a very simple sufficient condition, in the special case of two i.i.d.\ items, for the full-bundle mechanism (that just sets a single selling price for all the items together) to be optimal and deploy it to show that this holds for the equal-revenue distribution and, more generally, for all Pareto distributions with parameter $\alpha\geq 1/2$. Finally, \citet{Daskalakis:2013vn} have shown that full-bundling is also an optimal selling strategy for two exponentially i.i.d.\ goods. Nothing is clearly known in this front for more than three items or other distributions, apart from the recent work of \citet{gk2014} where a closed-form description of a deterministic mechanism that is shown to be optimal for up to six uniformly i.i.d.\ items is given, and conjectured that this holds for any number of goods. 

\emph{In the current paper we contribute to this line of work, by generalizing the result of~\citep{Daskalakis:2013vn} from $2$ to \emph{any number of goods}, showing that determinism is optimal for arbitrarily many identical valuations distributed according to an exponential distribution, and in fact optimality is achieved by the simple full-bundling strategy.} To our knowledge, this is the first result that provides specific description of an optimal auction for an arbitrary number of items. In fact, together with~\citep{gk2014}, they are the first such results to break the boundary of three items.

This path of discovering ``simple'' descriptions of optimal selling mechanisms is further being narrowed by a recent  computational hardness result from \citet{Daskalakis:2012fk}, where it is shown that even for independent (but not identical) valuations with finite support of size $2$, it is {\#}P-hard to compute exactly the  optimal mechanism. 
However this does not exclude the possibility of a PTAS and for i.i.d.\ settings \citet{Cai:2013kx} and \citet{Daskalakis:2012aa} have indeed already provided such efficient \emph{algorithmic} approximations.

So, given the previous discussion, it is essential to try to \emph{approximate} the optimal revenue by selling mechanisms that are as \emph{simple} as possible. We may lose something with respect to the total revenue objective, but on the other hand these mechanisms are much easier to understand, describe, analyze and implement, and such results may in fact enrich our understanding of the character of exact optimal auctions in general. \citet{Hart:2012uq} provide such solid and elegant approximation ratio guarantees (logarithmic with respect to the number of items) that hold ``universally'' for all product (independent) distributions, without even assuming standard regularity conditions (like e.g.\ in~\citep{Chawla:2007aa,Manelli:2006vn,Myerson:1981aa}), by studying the two most natural deterministic mechanisms: the one that treats every item independently and sells them separately to the buyer and, on the other end, the one that treats all items as a single full bundle. In particular, they prove an $O(\log^2 m)$-approximation ratio for the former and, with the extra requirement of identical distributions, an $O(\log m)$-approximation for the latter, where $m$ is the number of goods. They also provide slightly improved guarantees for the special case of two i.i.d.\ goods. \citet{Li:2013ty} further improved their results, by bringing these down to $\varTheta(\log m)$ and $\varTheta(1)$ respectively, which are also proved to be tight up to constants\footnote{In a very exciting recent result, after our paper was first made publicly available, \citet{Babaioff:2014ys} showed that combining these two simple selling mechanisms one can guarantee a constant approximation ratio of $6$ by just assuming independence of the item valuations. \citet{Yao:2014vn} generalized their results to multi-bidder settings.}.

\emph{In this paper we try to specialize these general probabilistic results for the case of specific ``canonical'' distributions, namely the uniform and exponential ones, and we show that by doing so one can get good constant-factor, almost optimal in many cases, performance guarantees.} Furthermore, in keeping up with the spirit of the line of work that studies simple but still well-performing mechanisms \emph{we also propose a very natural \emph{randomized} mechanism for exponential distributions and provide good approximation guarantees for it.}

\citet{Daskalakis:2013vn} and \citet{gk2014} have developed duality-theory frameworks for the general problem of multidimensional optimal auctions, the former having a strong measure-theoretic flavor by using classic results from optimal transport theory combined with Strassen's theorem for stochastic dominance, and the latter resembling more in spirit traditional linear programming theory formulations.

\subsection{Our Results and Techniques}

Our model is the standard single-buyer multi-item additive valuations Bayesian model of \citet{McAfee:1988nx} which is used in many other works~\citep{Manelli:2006vn,Hart:2012uq,Daskalakis:2013vn,Li:2013ty}. Critical to the exposition is the analytical characterization of truthfulness through subgradients of convex functions given by \citet{Rochet:1985aa}.

Inspired by the elegant approach of \citet{Hart:2012uq}, we take the opposite direction to their universal approximation guarantees for general independent distributions, and try to give better, specialized bounds for the case of uniform and exponential distributions. Our main strategy is driven by the standard technique in approximation algorithms, to use weak-duality from  traditional linear programming to upper-bound the optimal objective and then use this to calculate approximation upper bounds for particular algorithms. Since the optimal revenue problem cannot be fully captured by means of traditional combinatorial LPs, we use the duality-theory framework developed in \citep{gk2014}. In particular, we use the weak-duality theorem (see Theorem~\ref{th:weakduality}) in order to get specific closed-form bounds for our settings (Theorems~\ref{th:weakdualityuniform} and~\ref{th:weakdualityexpo}), by constructing and plugging-in appropriate feasible dual solutions (Theorems~\ref{th:optimaldualbounduniform} and \ref{th:exppbound}). This is the most technical part of the paper.
This technique is completely different to previous results on approximate mechanisms for the problem which rely entirely on probabilistic analysis methods (e.g.\ the core-tail decomposition of~\cite{Li:2013ty}).
Our bounds on the optimal revenue are very simple expressions, depending on the number of items $m$. We believe that, given how notoriously difficult is the problem of \emph{exactly} determining the optimal revenue, coming up with such formulas is a very useful tool for (approximate) auction analysis, and is of its own interest. 

By comparing these bounds to the revenue obtained by the simple mechanisms studied in \citet{Hart:2012uq} we are able to give closed-form approximation guarantees with respect to the number of items $m$, in both settings that we are interested in: for the case of i.i.d.\ uniform distributions (see Fig.~\ref{fig:unibundleratio}) over the unit interval we show that selling the items separately is loosely $2$-approximate and that selling in a full-bundle \emph{always} performs better and is asymptotically optimal; for independent (and not necessarily identical) exponential distributions (see Fig.~\ref{fig:exposepratio}) we give a closed-form formula upper bound~\eqref{eq:expsepratio} for selling separately that can be loosely upper-bounded by $e\approx 2.7$.

Furthermore, if the exponential distributions are in addition identical, then we can show (Theorem~\ref{eq:expoiidbundleoptimal}) that selling deterministically in a full-bundle is optimal, for \emph{any} number of goods.
We derive this optimality as a side result of the analysis of a very simple and natural randomized selling mechanism that we propose for the setting of independent (but not necessarily identical) exponential valuations. We call it \algoname{Proportional} (Definition~\ref{def:proportional}) and allocates the items somehow proportionally with respect to every item's exponential distribution parameter. We compute the expected revenue of this mechanism (Theorem~\ref{th:proportionalbound}) and using again the optimal revenue bounds derived earlier, we show that \algoname{Proportional}'s approximation ratio is at most equal to the ratio between the maximum and minimum parameters of the independent exponential distributions. For i.i.d.~settings, this ratio is of course equal to $1$, proving optimality and \algoname{Proportional} reduces to full-bundling.

A final remark must be made about the choice of uniform and exponential distributions. This was not random; we wanted to study ``canonical'' examples of distributions, one for bounded-interval supports and one having full support $[0,\infty)$. The uniform and exponential distributions are, respectively, the \emph{maximum entropy} probability distributions for these two settings, intuitively being the ``natural'' choices if one wants to make as few assumptions as possible (see e.g.~\citep[Sect.~3.4.3]{Golan:1996aa}).

\subsection{Model and Notation}\label{sec:model}

The real unit interval will be denoted by $I=[0,1]$ and the nonnegative reals by $\R_+=[0,\infty)$. We will also use $[m]=\ssets{1,2,\dots,m}$ for any positive integer $m$. The uniform distribution over $I$ will be denoted by $\mathcal U$ and the exponential distribution with parameter $\lambda >0$ by $\mathcal E(\lambda)$.  
Finally, we use the standard game-theoretic notation $\vecc x_{-j}=(x_1,x_2\dots,x_{j-1},x_{j+1},\dots,x_m)$ to denote the resulting vector if we remove $\vecc x$'s $j$-th coordinate. Then, $\vecc x=(x_j,\vecc x_{-j})$.

\subsubsection{Mechanisms and Truthfulness}

We study an $m$-goods monopoly setting, where the buyer has a valuation of $x_j\in D_j$ for good $j$. Every $D_j=[L_j,H_j]$ is a real interval. This valuation is private information to her, and intuitively represents the amount of money she is willing to pay to get this item. The seller has only some (incomplete) prior knowledge of the player's valuation vector $\vecc x=(x_1,x_2,\dots,x_m)$ in the form of a probability distribution $F$ over $D=D_1\times\dots\times D_m$ from which $\vecc x$ is sampled. For the purposes of this paper we will assume that $F=F_1\times\dots F_m$ is a product distribution, i.e.~$x_j$'s follow \emph{independent} distributions $F_j$, and also that every $F_j$ has an almost everywhere (a.e.) differentiable density function $f_j$.

A (direct revelation) selling mechanism $\mathcal M=(\vecc a,p)$ on this setting is a protocol which, after receiving the buyer's bid vector $\vecc x'$ as input (the buyer may lie about her true valuations $\vecc x$ and misreport $\vecc x'\neq\vecc x$), decides to sell item $j$  with probability $a_j(\vecc x')\in[0,1]$, $j\in[m]$, for a total (for all items) price of $p(\vecc x')\in\R_+$. If one wants to restrict attention only to deterministic mechanisms, it is enough to take $a_j\in\sset{0,1}$. However, here we optimize over the general class of randomized selling mechanisms, i.e.\ we allow lotteries.  

More formally, a mechanism consists of an \emph{allocation} function $\vecc a=(a_1,a_2,\dots,a_m):D\map I^m$ paired with a \emph{payment} function $p:D\map\R_+$. We consider the buyer having \emph{additive} valuations for the items, her ``happiness'' when she has (true) valuations $\vecc x$ and reports $\vecc x'$ to the mechanism being captured by her \emph{utility} function 
\begin{equation*}
u(\vecc x'|\vecc x)\equiv\vecc a(\vecc x')\cdot\vecc x-p(\vecc x')=\sum_{j=1}^ma_j(\vecc x')x_j-p(\vecc x'), 
\end{equation*}
the sum of the value she receives from the items she manages to purchase minus the payment she has to submit to the seller for this purchase. The player is completely rational and selfish, wanting to maximize her utility and that's why she will not hesitate to misreport $\vecc x'$ instead of her private values $\vecc x$ if this is to give her a higher utility. The seller's ``happiness'' is captured by the total \emph{revenue} of the mechanism 
\begin{equation}\label{eq:revenue}
p(\vecc x')=\vecc a(\vecc x')\cdot \vecc x-u(\vecc x'|\vecc x).
\end{equation}

It is standard in Mechanism Design to ask for mechanisms to respect the following two properties\footnote{We must mention here that these assumptions will be without loss to our revenue maximization objective, due to the celebrated Revelation Principle (see e.g.~\cite{Myerson:1981aa}). However, they help as simplify substantially the search for optimal selling mechanisms by narrowing the focus to truthful ones, under which we don't have to worry about the players' complex strategic behavior and the underlying game-theoretic equilibria.}:
\begin{itemize}

\item Individual Rationality (\textbf{IR}): $u(\vecc x|\vecc x)\geq 0$ for all $\vecc x\in D$.

\item Incentive Compatibility (\textbf{IC}): $u(\vecc x|\vecc x)\geq u(\vecc x'|\vecc x)$ for all $\vecc x,\vecc x'\in D$.

\end{itemize}
The IR constraint corresponds to the notion of voluntary participation, that is, the buyer cannot harm herself by taking part in the auction, while IC captures the fundamental property that truth-telling is a dominant strategy for the buyer in the underlying game, i.e.~she will never receive a better utility by lying about her true bid. Mechanisms that satisfy IC are also called \emph{truthful}. From now on we will focus on truthful IR mechanisms, and so we will relax notation $u(\vecc x'|\vecc x)$ to just $u(\vecc x)$, considering buyer's utility as a function $u:D\map\R_+$. 

\citet{Rochet:1985aa} gave a very elegant analytic characterization of truthful mechanisms (see e.g. \citep{Hart:2012uq} for a proof): $u$ must be a \emph{convex} function with the allocation function $\vecc a$ being its subgradient; $\frac{\partial u(\vecc x)}{\partial x_j}=a_j(\vecc x)$ for all items $j\in [m]$ and a.e.\ $\vecc x\in D$.
This essentially establishes a correspondence between truthful mechanisms and utility functions. Not only does every selling mechanism induce a well-defined utility function for the buyer, but also conversely, given a nonnegative convex function that satisfies these two properties we can fully recover a corresponding  mechanism from the allocation-derivative equality and~\eqref{eq:revenue}.

\subsubsection{Optimal Selling Mechanisms}\label{sec:optintro}

This paper studies the problem of maximizing the \emph{expected} seller's revenue based on his prior knowledge of distribution $ F$, given the IR and IC constraints, thus (by~\eqref{eq:revenue} and the subgradient characterization) maximizing
\begin{equation}\label{eq:totalrevenue}
\expect_{\vecc x\sim F }[p(\vecc x)] =\int_{D}\left(\nabla u(\vecc x)\cdot \vecc x - u(\vecc x)\right)\, d F(\vecc x)
\end{equation} 
over the space of nonnegative convex functions $u$ on $D$ having the property
\begin{equation*}\label{eq:diffs01}
0\leq \frac{\partial u(\vecc x)}{\partial x_j}\leq 1
\end{equation*}
for a.e.~$\vecc x\in D$ and all $j\in[m]$. 

Following the notation in~\citep{Hart:2012uq}, we will denote by $R(u; F)$ the expected revenue in~\eqref{eq:totalrevenue}. The \emph{optimal} revenue is $\algoname{Rev}( F)\equiv\sup_{u}R(u; F)$, i.e.~the maximum revenue among all feasible truthful mechanisms. A mechanism inducing utility $u$ for which $R(u; F)=\algoname{Rev}(F)$ will be called an \emph{optimal mechanism}. 

Notice that for single-dimensional $ F$'s, i.e.\  settings with a single good,  the seminal work of Myerson~\citep{Myerson:1981aa} has fully settled the question of determining optimal auctions and the optimal revenue (see e.g.~\citep{Hartline:2007aa}): 
\begin{equation}\label{eq:optimalsingle}
\algoname{Rev}(F)=\max_{p}p(1-F(p))
\end{equation}
obtained by the deterministic mechanism that sets a selling (take-it-or-leave-it) price threshold of $p$ ($p=\argmax_{p}p(1-F(p))$).  For example, if $F\sim U$ then $F(x)=x$ and $f(x)=1$ and so  by~\eqref{eq:optimalsingle} we can easily compute $\algoname{Rev}(U)=\frac{1}{4}$, achieved by the optimal mechanism that sets a selling price of $\frac{1}{2}$. Also, for the exponential distribution $\mathcal E(\lambda)$ we have $F(x)=1-e^{-\lambda x}$ and $f(x)=\lambda e^{-\lambda x}$ for $x\in\R_+$ and so $\algoname{Rev}(\mathcal E(\lambda))=\frac{1}{\lambda e}$ for a selling price of $\frac{1}{\lambda}$.
 
Finally, let us also provide some notation from~\citep{Hart:2012uq} for the revenue of two simple deterministic selling mechanisms for the $m$-items setting which are of great importance to us throughout this paper, namely the one which sells every item \emph{separately} to the buyer and that which sells all items in a \emph{full bundle}:
\begin{align}
\algoname{SRev}(F) &\equiv\algoname{Rev}(F_1)+\algoname{Rev}(F_2)+\dots+\algoname{Rev}(F_m)\label{eq:srevsum}\\
\algoname{BRev}(F) &\equiv\algoname{Rev}(F_1*F_2*\dots *F_m),\notag
\end{align}
respectively\footnote{$F_1*F_2*\dots *F_m$ is the \emph{convolution} of distributions $F_1,F_2,\dots,F_m$, i.e.~if $X_j\sim F_j$ are the random variables representing the items' valuations, then $F_1*F_2*\dots *F_m=F_S$ is the distribution of the sum of the valuations, $S=\sum_{j=1}^mX_j$.}. In particular, if $F_j\sim U$ for all $j\in[m]$, we have 
\begin{equation}\label{eq:sepandbundleuniform}
\algoname{SRev}(U^m)=\frac{m}{4}\quad\text{and}\quad \algoname{BRev}(U^m)=\sup_{x\in[0,m]} x(1-F_S(x)),
\end{equation}
where $F_S$ is the cdf of the Irwin-Hall distribution of the sum of $m$ independent uniform random variables over $I$, i.e.~\citep{Hall:1927aa}
\begin{equation}\label{eq:irwinhallcdf}
F_S(x)=\frac{1}{m!}\sum_{k=0}^{\lfloor x\rfloor}(-1)^k\binom{m}{k}(x-k)^m,\quad 0\leq x\leq m.
\end{equation}
In the same way, for independent exponential distributions and for i.i.d.~exponential distributions, we can see that
\begin{equation}\label{eq:expsep}
\algoname{SRev}(\mathcal E)=\frac{1}{e}\sum_{j=1}^m\frac{1}{\lambda_j}\quad\text{and}\quad\algoname{SRev}(\mathcal E^m(\lambda))=\frac{m}{\lambda e},
\end{equation}
where we use the shorthand notation $\mathcal E=\mathcal E(\lambda_1)\times\dots\times\mathcal E(\lambda_m)$.

\section{Weak Duality}\label{sec:weakduality}

Define $\mathcal V_{D,F}=\mathcal V$ to be the family of all $m$-dimensional functions $\vecc z=(z_1,z_2,\dots,z_m):D\map\R_+^m$, such that every $z_j$ is integrable and also absolutely continuous with respect to its $j$-th coordinate, and which satisfy the following conditions a.e.\ in $D$:
\begin{equation}
z_j(L_j,\vecc x_{-j}) \leq L_jf(L_j,\vecc x_{-j})\quad\text{for all}\;\; j\in[m]\label{eq:duals0}
\end{equation}
\begin{equation}
z_j(H_j,\vecc x_{-j}) \geq H_jf(H_j,\vecc x_{-j})\quad\text{for all}\;\; j\in[m]\label{eq:duals1} 
\end{equation}
\begin{equation} 
\sum_{j=1}^m\frac{\partial z_j(\vecc x)}{\partial x_j}\leq (m+1)f(\vecc x)+\vecc x\cdot\nabla f(\vecc x).\label{eq:boundedderivdual}
\end{equation}
Then, we know from the duality framework developed in \citep{gk2014} that the optimal revenue can be bounded by the following quantity:

\begin{theorem}[Weak Duality]
\label{th:weakduality} 
For every $m$-dimensional interval $D$ and probability distribution $F$ over $D$,
\begin{equation}
\revenue{F}\leq\inf_{\vecc z\in\mathcal V}\int_{D}\sum_{j=1}^mz_j(\vecc x)\,d\vecc{x}.
\label{eq:dualobjective,eq:dualprob}
\end{equation}
\end{theorem}
We will feel free to refer to $\vecc z\in\mathcal V$ as feasible \emph{dual solutions}, constraints \eqref{eq:duals0}--\eqref{eq:boundedderivdual} as \emph{dual constraints} and to the above critical quantity $\inf_{\vecc z\in\mathcal V}\int_{D}\sum_{j} z_j(\vecc x)\,d\vecc{x}$ as \emph{dual objective}. The ties to the terminology of classical linear programming duality are obvious. 

Since we are particularly interested in providing upper bounds for the optimal revenue in the case of independent uniform and exponential valuations, we specialize Theorem~\ref{th:weakduality} in these cases, for ease of reference and a more clear exposition:
\begin{theorem}[Weak Duality for Uniform Distributions]\label{th:weakdualityuniform}
The dual constraints \eqref{eq:duals0}--\eqref{eq:boundedderivdual} for the $m$-items uniform i.i.d.\ setting over $I^m$ become:
\begin{equation}
z_j(0,\vecc x_{-j}) = 0\quad\text{for all}\;\; j\in[m]\label{eq:duals0uniform}
\end{equation}
\begin{equation}
z_j(1,\vecc x_{-j}) \geq 1\quad\text{for all}\;\; j\in[m]\label{eq:duals1uniform} 
\end{equation}
\begin{equation} 
\sum_{j=1}^m\frac{\partial z_j(\vecc x)}{\partial x_j}\leq m+1,\label{eq:boundedderivdualuniform}
\end{equation}
and the dual objective upper-bounds optimal revenue:
\begin{equation}
\revenue{\mathcal U^m}\leq\sum_{j=1}^m\int_{I^m} z_j(\vecc x)\,d\vecc{x}
\label{eq:dualobjectiveuniform}.
\end{equation}
\end{theorem}
Intuitively, for each item $j$ there is a dual function $z_j$ that, if viewed as a single dimensional function $z_j(*,\vecc x_{-j})$ of just the $j$-th coordinate, it has to start with the value of $0$ at the left boundary of the unit interval $I$ (condition~\eqref{eq:duals0uniform}), travel all the way up to a value of $1$ at the end of the interval (condition~\eqref{eq:duals1uniform}), but with a speed that is never too high: the sum of the derivatives of all $z_j$'s must not exceed $m+1$ at any point of the domain (condition~\eqref{eq:boundedderivdualuniform}). At the same time, in order to get good bounds on the optimal revenue, these functions must stay as low as possible, so that they minimize the total volume under their curves (expression~\eqref{eq:dualobjectiveuniform}). These are two contradicting objectives, and finding the right balance between them is the very essence of this duality method. 

\begin{theorem}[Weak Duality for Exponential Distributions]\label{th:weakdualityexpo}
The dual constraints \eqref{eq:duals0}--\eqref{eq:boundedderivdual} for the $m$-items independent exponential setting (with parameters $\lambda_1,\lambda_2,\dots,\lambda_m$) become:
\begin{equation}
z_j(0,\vecc x_{-j}) = 0\quad\text{for all}\;\; j=1,2,\dots,m\label{eq:duals0expo}
\end{equation}
\begin{equation} 
\sum_{j=1}^m\frac{\partial z_j(\vecc x)}{\partial x_j}\leq\lambda\left(m+1- w\right)e^{- w},\label{eq:boundedderivdualexpo}
\end{equation}
where $w=\sum_{j=1}^m\lambda_jx_j$, $\lambda=\prod_{j=1}^m\lambda_j$ and the dual objective upper-bounds optimal revenue:
\begin{equation}
\revenue{\mathcal E}\leq\sum_{j=1}^m\int_{\R^m_{+}}z_j(\vecc x)\,d\vecc{x}
\label{eq:dualobjectiveexpo}.
\end{equation}
\end{theorem}

\section{Uniform Domains}\label{sec:uniform}
\begin{figure}
\begin{center}
\begin{tikzpicture}[domain=0:100,xscale=0.09,yscale=0.09]
\draw[dotted,step=5,very thin,color=gray] (0,0) grid (105,55);
  \draw[->] (0,0) -- (110,0) node[below] {$m$};
  \draw[->] (0,0) -- (0,55);
\draw[thick,color=blue] (2.,5.55556) -- (3.,12.5) -- (4.,18.) -- (5.,22.2222) -- (6.,25.5102) -- (7.,28.125) -- (8.,30.2469) -- (9.,32.) -- (10.,33.4711) -- (11.,34.7222) -- (12.,35.7988) -- (13.,36.7347) -- (14.,37.5556) -- (15.,38.2813) -- (16.,38.9273) -- (17.,39.5062) -- (18.,40.0277) -- (19.,40.5) -- (20.,40.9297) -- (21.,41.3223) -- (22.,41.6824) -- (23.,42.0139) -- (24.,42.32) -- (25.,42.6036) -- (26.,42.8669) -- (27.,43.1122) -- (28.,43.3413) -- (29.,43.5556) -- (30.,43.7565) -- (31.,43.9453) -- (32.,44.123) -- (33.,44.2907) -- (34.,44.449) -- (35.,44.5988) -- (36.,44.7407) -- (37.,44.8753) -- (38.,45.0033) -- (39.,45.125) -- (40.,45.2409) -- (41.,45.3515) -- (42.,45.457) -- (43.,45.5579) -- (44.,45.6543) -- (45.,45.7467) -- (46.,45.8352) -- (47.,45.9201) -- (48.,46.0017) -- (49.,46.08) -- (50.,46.1553) -- (51.,46.2278) -- (52.,46.2976) -- (53.,46.3649) -- (54.,46.4298) -- (55.,46.4923) -- (56.,46.5528) -- (57.,46.6112) -- (58.,46.6676) -- (59.,46.7222) -- (60.,46.7751) -- (61.,46.8262) -- (62.,46.8758) -- (63.,46.9238) -- (64.,46.9704) -- (65.,47.0156) -- (66.,47.0595) -- (67.,47.1021) -- (68.,47.1435) -- (69.,47.1837) -- (70.,47.2228) -- (71.,47.2608) -- (72.,47.2978) -- (73.,47.3338) -- (74.,47.3689) -- (75.,47.403) -- (76.,47.4363) -- (77.,47.4688) -- (78.,47.5004) -- (79.,47.5313) -- (80.,47.5613) -- (81.,47.5907) -- (82.,47.6194) -- (83.,47.6474) -- (84.,47.6747) -- (85.,47.7015) -- (86.,47.7276) -- (87.,47.7531) -- (88.,47.7781) -- (89.,47.8025) -- (90.,47.8263) -- (91.,47.8497) -- (92.,47.8726) -- (93.,47.895) -- (94.,47.9169) -- (95.,47.9384) -- (96.,47.9594) -- (97.,47.98) -- (98.,48.0002) -- (99.,48.02) -- (100.,48.0394);
\node[right] at (100.,48.0394) {$\frac{\algoname{Rev}(\mathcal U^m)}{\algoname{SRev}(\mathcal U^m)}$};
\draw[thick,color=red] (2,0.475373) -- (3,3.75) -- (4,6.09055) -- (5,7.64896) -- (6,8.6866) -- (7,9.38345) -- (8,9.85351) -- (9,10.1692) -- (10,10.3774) -- (11,10.5094) -- (12,10.5865) -- (13,10.6236) -- (14,10.6311) -- (15,10.6166) -- (16,10.5858) -- (17,10.5427) -- (18,10.4904) -- (19,10.4313) -- (20,10.3672) -- (21,10.2995) -- (22,10.2292) -- (23,10.1572) -- (24,10.0842) -- (25,10.0106) -- (26,9.93682) -- (27,9.86322) -- (28,9.79003) -- (29,9.71745) -- (30,9.6456) -- (31,9.57462) -- (32,9.50459) -- (33,9.43557) -- (34,9.36761) -- (35,9.30075) -- (36,9.23499) -- (37,9.17035) -- (38,9.10685) -- (39,9.04446) -- (40,8.9832) -- (41,8.92304) -- (42,8.86397) -- (43,8.80598) -- (44,8.74905) -- (45,8.69316) -- (46,8.63829) -- (47,8.58441) -- (48,8.53151) -- (49,8.47957) -- (50,8.42855) -- (51,8.37845) -- (52,8.32924) -- (53,8.28089) -- (54,8.2334) -- (55,8.18672) -- (56,8.14085) -- (57,8.09577) -- (58,8.05145) -- (59,8.00788) -- (60,7.96503) -- (61,7.92289) -- (62,7.88145) -- (63,7.84068) -- (64,7.80056) -- (65,7.76109) -- (66,7.72224) -- (67,7.68401) -- (68,7.64636) -- (69,7.6093) -- (70,7.5728) -- (71,7.53686) -- (72,7.50145) -- (73,7.46658) -- (74,7.43221) -- (75,7.39835) -- (76,7.36498) -- (77,7.33208) -- (78,7.29966) -- (79,7.26769) -- (80,7.23616) -- (81,7.20508) -- (82,7.17442) -- (83,7.14418) -- (84,7.11434) -- (85,7.08491) -- (86,7.05586) -- (87,7.0272) -- (88,6.99891) -- (89,6.97098) -- (90,6.94342) -- (91,6.9162) -- (92,6.88933) -- (93,6.86279) -- (94,6.83658) -- (95,6.8107) -- (96,6.78513) -- (97,6.75987) -- (98,6.73491) -- (99,6.71025) -- (100,6.68588);
\node[right] at ((100,6.68588) {$\frac{\algoname{Rev}(\mathcal U^m)}{\algoname{BRev}(\mathcal U^m)}$};
 \draw[dashed,color=gray] (0,50) -- (105,50);
 \draw[dashed,color=gray] (100,0) -- (100,55.0002);
 \node[below] at (50,0) {\scriptsize$50$};
 \node[below] at (75,0) {\scriptsize$75$};
 \node[below] at (100,0) {\scriptsize$100$};
 \node[below] at (25,0) {\scriptsize$25$};
 \node[left] at (0,50) {\scriptsize$2$};
  \node[left] at (0,5) {\scriptsize$1.1$};
  \node[left] at (0,20) {\scriptsize$1.4$};
  \node[left] at (0,35) {\scriptsize$1.7$};
\end{tikzpicture}
\end{center}
\caption{The approximation ratio bounds for the uniform i.i.d.~separate and full-bundle selling mechanisms in~\eqref{eq:unisepratio} and~\eqref{eq:unibundleratio}.}
\label{fig:unibundleratio}
\end{figure}
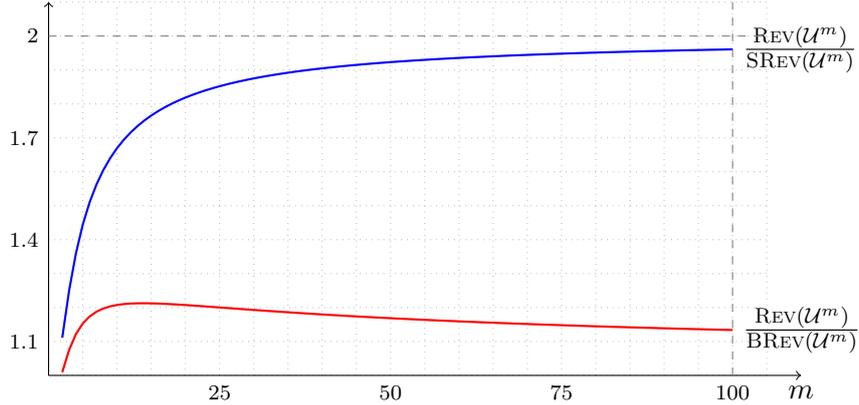

\begin{mdframed}
\begin{theorem}\label{th:optimaldualbounduniform}
The optimal revenue from selling $m$ goods having uniform i.i.d.~valuations over the unit interval is at most
$$
\frac{m(1+m^2)}{2(1+m)^2}.
$$ 
\end{theorem}
\end{mdframed}
\begin{proof}
Let  $$\mathcal I_m\equiv\sset{\vecc v=(v_1,v_2,\dots,v_m)\fwh{v_j\in\sset{0,1}, j\in[m]}}\label{page:unithyper}$$
be the set of nodes of the $m$-dimensional unit hypercube and for every node $\vecc v\in\mathcal I_m$ define $L_{\vecc v}$ to be the following subspace of $I^m$:
$$
L_{\vecc v}=\sset{\vecc x\in I^m\fwh{x_j\in\left[0,\frac{1}{m+1}\right]\;\;\text{if}\;\; v_j=0\;\;\text{and}\;\;x_j\in\left(\frac{1}{m+1},1\right]\;\;\text{if}\;\; v_j=1,\quad j\in[m]}}
$$
A simple observation is that $L_\vecc{v}$'s form a valid partition of $I^m$, i.e.
$$
 \vecc v,\vecc v'\in\mathcal I_m \land \vecc v\neq \vecc v'\then L_\vecc{v}\inters L_{\vecc v'}=\emptyset\qquad\text{and}\qquad\bigunion_{\vecc v\in \mathcal I_m}L_{\vecc v}=I^m
$$

Now we are going to construct a feasible dual solution, that is, valid $z_j$'s, to plug them into Theorem \ref{th:weakdualityuniform}. Fix some $j\in[m]$ and a subspace $L_{\vecc v}\subseteq I^m$ (by fixing a $\vecc v=(v_1,v_2,\dots,v_m)\in \mathcal I_m$) and define $z_j:L_{\vecc v}\map\R_{+}$ as follows:
\begin{itemize}
\item If $v_j=0$, set $z_j(\vecc x)=0$ for all $\vecc x\in L_{\vecc v}$.
\item Otherwise, i.e.~if $v_j=1$, set
$$
z_j(\vecc x)=\max\sset{0,\frac{m+1}{k}(x_j-c_k)}=\begin{cases}0,&\text{if}\;\; \frac{1}{m+1}<x_j\leq c_k,\\ \frac{m+1}{k}(x_j-c_k),&\text{if}\;\; c_k<x_j\leq 1, \end{cases}
$$
for all $\vecc x\in L_{\vecc v}$, where 
$$k=k(\vecc v)=\sum_{j=1}^m v_j\quad\text{and}\quad c_k=1-\frac{k}{m+1}.$$
\end{itemize}
By this construction, and by letting $\vecc v$ range over $\mathcal I_m$, we have a well defined function $z_j:I^m\map\R_{+}$. Each $\vecc x\in[0,1]^m$ belongs to a unique partition $L_{\vecc v}$ (corresponding to a \emph{unique} $\vecc v=\vecc v(\vecc x)$), thus also well defining $k=k(\vecc x)$. So, the above definition can be written more compactly as
$$
z_j(\vecc x)=\begin{cases}0,&\text{if}\;\; 0<x_j\leq c_k,\\ \frac{m+1}{k}(x_j-c_k),&\text{if}\;\; c_k<x_j\leq 1. \end{cases}
$$

It is easy to check, directly from this definition, that
\begin{equation} \label{eq:zuniform01}
z_j(0,x_{-j})=0\quad\text{and}\quad z_j(1,x_{-j})=1
\end{equation}
for all $j=1,2,\dots,m$ and $x_{-j}\in I^{m-1}$ and also that  
\begin{equation}\label{eq:zderivs}
\frac{\partial z_j(\vecc x)}{\partial x_j}=\begin{cases}0,&\text{if}\;\; 0<x_j\leq c_k,\\ \frac{m+1}{k},&\text{if}\;\; c_k<x_j\leq 1. \end{cases}
\end{equation}

Furthermore, if we fix some $\vecc x\in I^m$ (and thus also fix the corresponding, well-defined, $\vecc v=\vecc v(\vecc x)\in \mathcal I_m$ and $k=\sum_{j=1}^m v_j$), we see from property \eqref{eq:zderivs} above that
\begin{equation}\label{eq:zuniformderivbound}
\sum_{j=1}^m\frac{\partial z_j(\vecc x)}{\partial x_j}\leq \sum_{j=1}^mv_j\frac{m+1}{k}=\frac{m+1}{k} \sum_{j=1}^mv_j=m+1.
\end{equation}
But now we can see that Eqs.~\eqref{eq:zuniform01} and \eqref{eq:zuniformderivbound} are exactly properties \eqref{eq:duals0uniform}, \eqref{eq:duals1uniform} and \eqref{eq:boundedderivdualuniform}.

The last remaining step of the proof is to evaluate the dual objective and show that $$\int_{I^m}\sum_{j=1}^mz_j(\vecc x)\,d\vecc{x}=\frac{m(1+m^2)}{2(1+m)^2},$$ which we do in Appendix~\ref{append:uniform1}.
\end{proof}
Theorem~\ref{th:optimaldualbounduniform} combined with~\eqref{eq:sepandbundleuniform} immediately gives us the following approximation ratio bound for the simple deterministic mechanism that sells the items separately:
\begin{equation}\label{eq:unisepratio}
\frac{\algoname{Rev}(\mathcal U^m)}{\algoname{SRev}(\mathcal U^m)}\leq 2\frac{1+m^2}{(1+m)^2}<2.
\end{equation}  
A plot of this approximation ratio for the values of $m=1,2,\dots,100$ can be found in Fig.~\ref{fig:unibundleratio}, drawn with blue color. 
In the same way, using the other expression of~\eqref{eq:sepandbundleuniform} together with~\eqref{eq:irwinhallcdf} we get a bound for the approximation ratio of the deterministic full bundle mechanism, which is asymptotically optimal:
\begin{equation}\label{eq:unibundleratio}
\frac{\algoname{Rev}(\mathcal U^m)}{\algoname{BRev}(\mathcal U^m)}\leq \frac{m(1+m^2)}{2(1+m)^2\sup_{x\in[0,m]}x\left(1-\frac{1}{m!}\sum_{k=0}^{\lfloor x\rfloor}(-1)^k\binom{m}{k}(x-k)^m\right)}\to 1,
\end{equation}
as $m\to\infty$. A plot of this approximation ratio for the values of $m=1,2,\dots,100$ can be found in Fig.~\ref{fig:unibundleratio}, drawn with red color. \emph{Notice how full bundling outperforms selling separately for \emph{any} number of goods $m$.}
\paragraph{Discussion of Theorem~\ref{th:optimaldualbounduniform}}
We must mention here that one can trivially get an upper bound of $\frac{m}{2}$ for the optimal revenue $\algoname{Rev}(\mathcal U^m)$, which is close to that given by Theorem~\ref{th:optimaldualbounduniform}: simply observe that by the IR constraint the seller cannot expect to extract more revenue than the buyer's surplus, i.e.\ the sum of the item valuations $\sum_{j=1}^m x_j$, which for the uniform distribution has an expectation of $\frac{m}{2}$. The two bounds are equal in the limit as the number of items grows large, however the one in Theorem~\ref{th:optimaldualbounduniform} still gives an improvement by a factor of $\frac{(m+1)^2}{m^2+1}$, which especially for a small number of goods, is not insignificant. Notice that, due to the Law of Large Numbers, the optimal revenue as $m\to\infty$ will anyway tend to the expected full surplus, not only for the uniform distribution, but for any kind of independently distributed items\footnote{For more on this, see the discussion in~\citep{Hart:2012uq}. This is the reason why, as~\citet{Hart:2014kl} state, for our problem ``the difficult case is when there is more than one but not too many goods''.}. 

\begin{figure}
\centering
\begin{subfigure}{0.95\textwidth}
\centering
\includegraphics[width=0.43\textwidth]{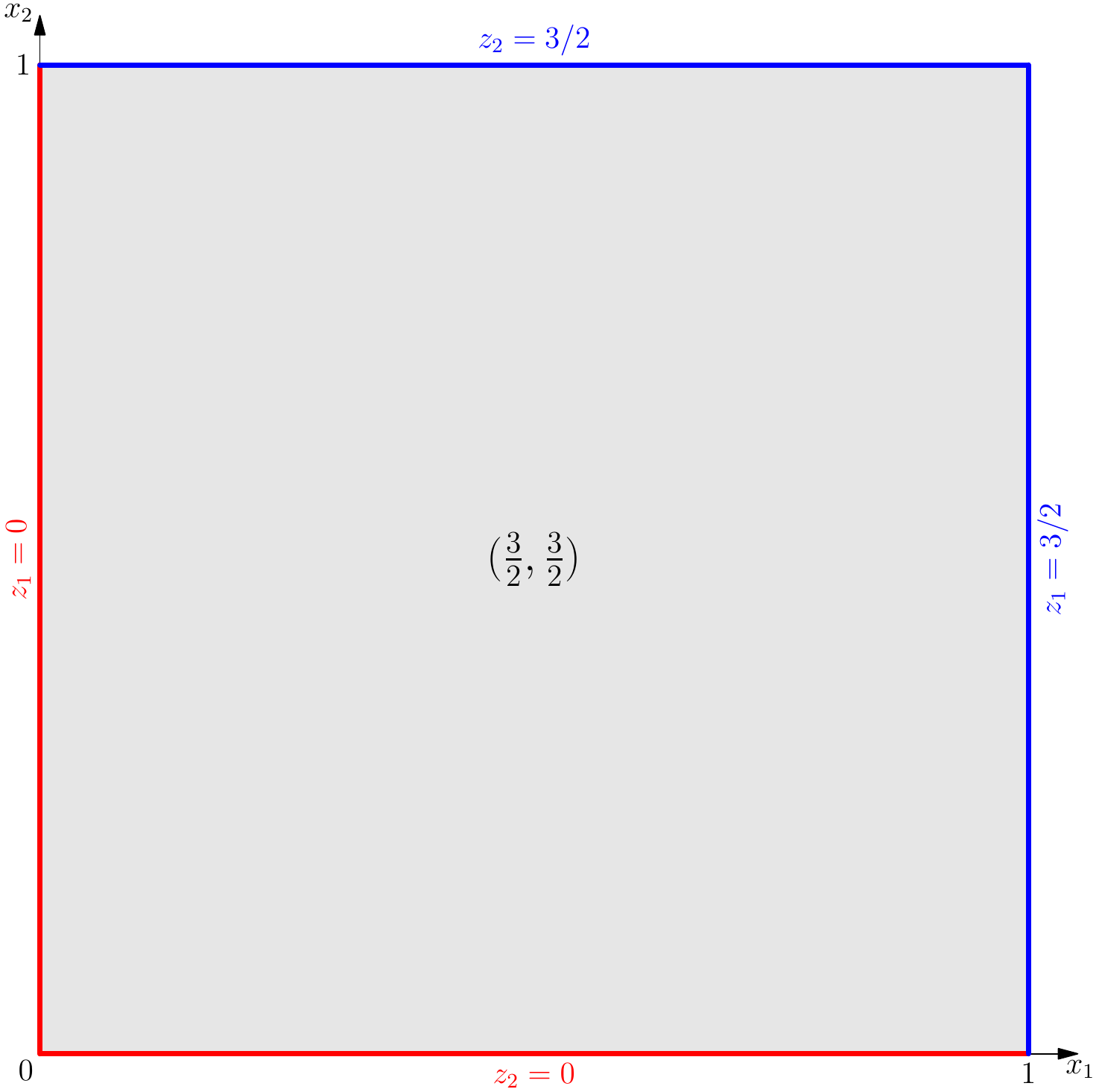}
\caption{\footnotesize Trivial solution}
\label{fig:dual_solutions_uniform_two_items_trivial}
\end{subfigure}
~
\begin{subfigure}{0.95\textwidth}
\centering
\includegraphics[width=0.43\textwidth]{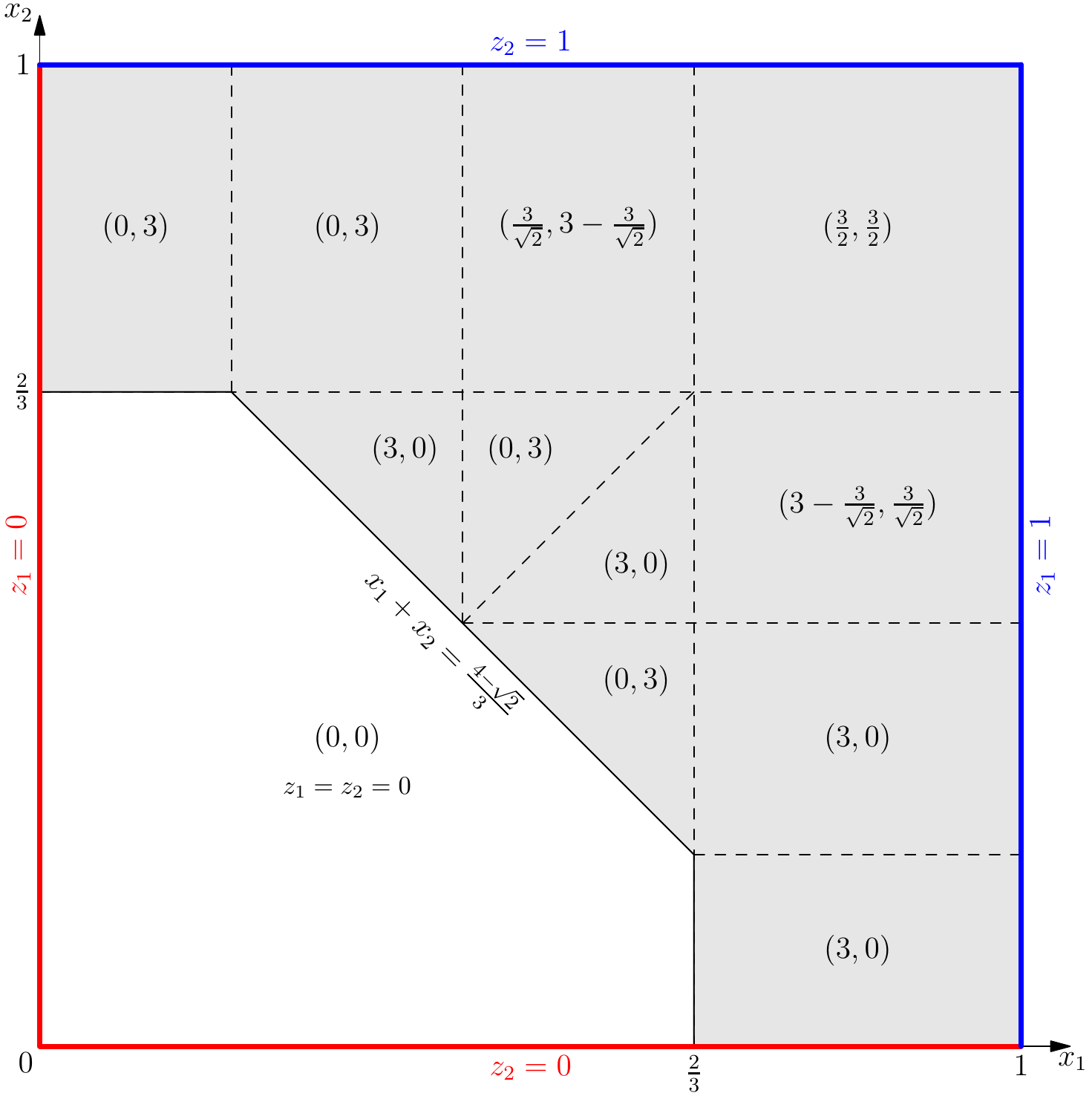}
\caption{\footnotesize Optimal solution}
\label{fig:dual_solutions_uniform_two_items_opt}
\end{subfigure}
~
\begin{subfigure}{0.95\textwidth}
\centering
\includegraphics[width=0.43\textwidth]{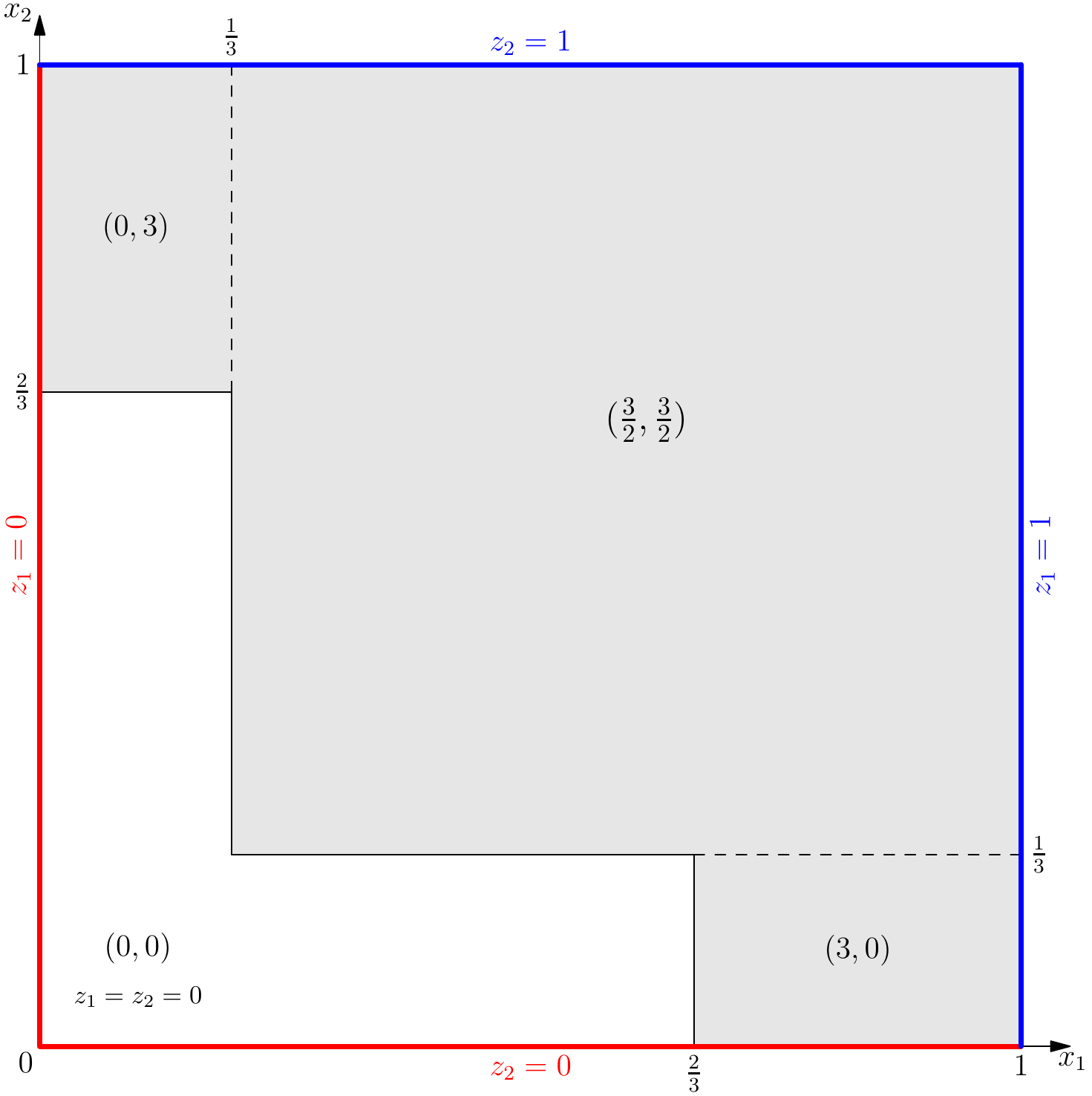}
\caption{\footnotesize The solution in Theorem~\ref{th:optimaldualbounduniform}}
\label{fig:dual_solutions_uniform_two_items_theorem}
\end{subfigure}
\caption{Different feasible dual solutions $z_1,z_2$ for two goods with uniform valuations over $[0,1]$, given by their critical derivatives $\left(\frac{\partial z_1(x_1,x_2)}{\partial x_1},\frac{\partial z_2(x_1,x_2)}{\partial x_2}\right)$.}
\label{fig:dual_solutions_uniform_two_items}
\end{figure}

From that perspective, we believe that it is interesting to get bounds on the optimal revenue other than the trivial ones derived from using the above surplus-bound argument. To our knowledge Theorem~\ref{th:optimaldualbounduniform} is the first such result in the literature. But, probably even more important than the improvement in the bound's value itself, is the underlying technique of providing \emph{explicit} feasible dual solutions to plug into Theorem~\ref{th:weakduality}: this can give new insights in the structure of good approximations of the optimal revenue, something which is known to be particularly difficult for our problem. To demonstrate this, consider for example the case of just two goods ($m=2$) with valuations drawn uniformly from $I$. The dual program \eqref{eq:duals0uniform}--\eqref{eq:boundedderivdualuniform} essentially asks to allocate a total available value of $3$ among the derivatives of $z_1$ and $z_2$ over $I^2$ in a way that these functions start at $0$ and grow up above $1$ at the boundary of $I$. On one hand there is the trivial way to do that, simply allocating the total weight equally: $\frac{\partial z_1(\vecc x)}{\partial x_1}=\frac{\partial z_2(\vecc x)}{\partial x_2}=\frac{3}{2}$ for all $\vecc x\in I^2$; this is clearly a suboptimal solution, since the functions end up reaching a value of $\frac{3}{2}$ at the boundary's end, way above $1$. On the other hand, there is the optimal way to do it, given in~\citep{g2014_2} (see Fig.~\ref{fig:dual_solutions_uniform_two_items_opt}). However, this construction is rather involved and very difficult to generalize, and in fact only \emph{existential} proofs of optimality are know for more goods, and only up to $6$ items. So, it seems essential to find some middle ground within these two extremes, providing a good approximation to the optimal revenue but also still being simple enough to generalize for any number of items $m$. This is exactly what the construction of the dual solutions in the proof of Theorem~\ref{th:optimaldualbounduniform} provides. A demonstration is given in Fig.~\ref{fig:dual_solutions_uniform_two_items} for the case of two goods.

\section{Exponential Domains}\label{sec:expoenential}
In the following, for $m$ positive integer and $w\in\R_+$ we will denote the (upper) incomplete Gamma function by
$$
\varGamma(m,w)\equiv\int_w^\infty t^{m-1}e^{-t}\,dt=(m-1)!e^{-w}\sum_{k=0}^{m-1}\frac{w^k}{k!}
$$
and define
\begin{equation}\label{eq:funcg}
g(m,w)=\varGamma(m+1,w)-(m+1)\varGamma(m,w).
\end{equation}
Function $g$ is continuous and has a unique root with respect to variable $w\in\R_+$. Let's denote this root by $\gamma^*_m$. In fact, $\gamma^*_m\in(0,m+1)$ and also $g(m,w)>0$ for all $w\in(\gamma^*_m,\infty)$. So, the following function on the positive integers is well defined and nonnegative:
\begin{equation}\label{eq:Gfunction}
G(m)\equiv \int_{0}^\infty \max\{0,g(m,w)\}\,dw =\int_{\gamma^*_m}^\infty g(m,w)\,dw=\gamma^*_m\varGamma(m,\gamma^*_m)={\gamma^*_m}^{m+1}e^{-\gamma^*_m}.
\end{equation}
A detailed proof of all the above properties of function $g(m,w)$ and calculations can be found in Appendix~\ref{append:funcg}.

\begin{mdframed}
\begin{theorem}\label{th:exppbound}
The optimal revenue from selling  $m$ goods having independent exponential valuations (with parameters $\lambda_1,\lambda_2,\dots,\lambda_m$) is at most
$$
\frac{G(m)}{m!}\sum_{j=1}^m\frac{1}{\lambda_j},
$$
where $G(m)$ is defined in~\eqref{eq:Gfunction}.
In the special case of i.i.d.\ exponential valuations with parameter $\lambda$ this becomes
$$
\frac{G(m)}{(m-1)!\lambda}.
$$
\end{theorem}
\end{mdframed}
\begin{proof}
We will construct appropriate dual variables $z_j$ that satisfy~\eqref{eq:duals0expo} and~\eqref{eq:boundedderivdualexpo} to plug into the Weak Duality Theorem~\ref{th:weakdualityexpo}. For all $j=1,2,\dots,m$ and $\vecc x\in\R^m_+$ we define
$$
z_j(\vecc x)=\max\sset{0,\hat\lambda x_jw^{-m}g(m,w)}=
\begin{cases}
\hat\lambda x_jw^{-m}g(m,w), &\text{if}\;\;w\geq \gamma^*_m\\
0, &\text{otherwise},
\end{cases}
$$
where $\hat\lambda=\prod_{j=1}^m\lambda_j$, $w=\sum_{j=1}^m\lambda_jx_j$ and function $g(m,w)$ as defined in~\eqref{eq:funcg}. The nonnegativity of the dual variables as well as their absolute continuity is immediate from the properties of function $g(m,w)$. It is also trivial to see that condition~\eqref{eq:duals0expo} is immediately satisfied by the definition of $z_j$. Regarding condition~\eqref{eq:boundedderivdualexpo}, for any $j\in [m]$ and $\vecc x\in\R_+^m$ such that $w\geq \gamma^*_m$ we calculate
\begin{align*}
\frac{\partial z_j(\vecc x)}{\partial x_j} &=\hat\lambda w^{-m}g(m,w)+\hat\lambda x_j\frac{\partial w^{-m}g(m,w)}{\partial x_j}\\
		&=\hat\lambda w^{-m}g(m,w)+\hat\lambda\lambda_j x_j\frac{\partial w^{-m}g(m,w)}{\partial w}\\
		&=\hat\lambda w^{-m}g(m,w)+\hat\lambda\lambda_j x_j\left[w^{-m}\frac{\partial g(m,w)}{\partial w}-mw^{-m-1}g(m,w)\right]\\
		&=\hat\lambda w^{-m}g(m,w)+\hat\lambda\lambda_j x_j\left[(m+1-w)w^{-1}e^{-w}-mw^{-m-1}g(m,w),\right]
\end{align*}
so, by summing up we get
\begin{align*}
\sum_{j=1}^m\frac{\partial z_j(\vecc x)}{\partial x_j} &=m\hat\lambda w^{-m}g(m,w)+\hat\lambda w\left[(m+1-w)w^{-1}e^{-w}-mw^{-m-1}g(m,w)\right]\\
		&=\hat\lambda(m+1-w)e^{-w}.
\end{align*}
At the remaining case of $w<\gamma^*_m$, we have that $z_j(\vecc x)=0$ for all $j\in[m]$. Also, since $\gamma^*_m<m+1$, we know that $w<m+1$, so
$$
\sum_{j=1}^m\frac{\partial z_j(\vecc x)}{\partial x_j}=0<\hat\lambda(m+1-w)e^{-w}.
$$
Thus, in any case~\eqref{eq:boundedderivdualexpo} is satisfied.

Finally, we compute the dual objective value in~\eqref{eq:dualobjectiveexpo}. First notice that 
$$
\sum_{j=1}^mz_j(\vecc x)=
\begin{cases}
\hat\lambda w^{-m}g(m,w)\sum_{j=1}^m x_j, &\text{if}\;\;w\geq \gamma^*_m\\
0, &\text{otherwise}.
\end{cases}
$$
We perform the following change of variables in the integral:
\begin{equation}\label{eq:changevar1}
x_1=t_1\frac{w}{\lambda_1},x_2=t_2\frac{w}{\lambda_2},\dots,x_{m-1}=t_{m-1}\frac{w}{\lambda_{m-1}}\quad\text{and}\quad x_m=(1-t_1-t_2-\dots-t_{m-1})\frac{w}{\lambda_m}
\end{equation}
where $w=\sum_{j=1}^m\lambda_jx_j\in\R_+$ and $t_1,\dots,t_{m-1}\in\R_+$ with $0\leq t_1+\dots+t_{m-1}\leq 1$.
Denote this subspace of $I^{m-1}$ where $t_j$'s range by $\mathcal A$. The Jacobian of this transformation equals $\frac{w^{m-1}}{\hat\lambda}$ and so the integral in~\eqref{eq:dualobjectiveexpo} can be written as:
\begin{align*}
\int_{\R^m_{+}}\sum_{j=1}^mz_j(\vecc x)\,d\vecc{x} &=\int_{\gamma^*_m}^{\infty}\int_{\mathcal A}\lambda w^{-m}g(m,w)\sum_{j=1}^m x_j\cdot\frac{w^{m-1}}{\hat\lambda}\,dt_{1}\,dt_2\dots\,dt_{m-1}\,dw\\
		&=\int_{\gamma^*_m}^{\infty}g(m,w)\,dw\int_\mathcal{A}\frac{t_1}{\lambda_1}+\dots+\frac{t_{m-1}}{\lambda_{m-1}}+\frac{1-t_1-\dots-t_{m-1}}{\lambda_m}\,dt_{1}\dots\,dt_{m-1}\\
		&=G(m)\int_\mathcal{A}\sum_{j=1}^{m-1}\left(\frac{1}{\lambda_j}-\frac{1}{\lambda_m}\right)t_{j}+\frac{1}{\lambda_m}\,dt_{1}\dots\,dt_{m-1}\\
		&=G(m)\left[\frac{1}{m!}\sum_{j=1}^{m-1}\left(\frac{1}{\lambda_j}-\frac{1}{\lambda_m}\right)+\frac{1}{(m-1)!\lambda_m}\right]\\
		&=\frac{G(m)}{m!}\sum_{j=1}^m\frac{1}{\lambda_j}.
\end{align*} 
At the third equation above we used the following Lemma describing some known ``geometric'' properties of the body $\mathcal A$ used in the transformation~\eqref{eq:changevar1}. 
\begin{lemma}\label{lemma:chainunit}
For any positive integer $m$,
\begin{equation*}
\int_\mathcal{A}1\,dt_1\dots\,dt_{m-1}=\mu(\mathcal A)=\frac{1}{(m-1)!}\quad\text{and}\quad\int_\mathcal{A}t_j\,dt_1\dots\,dt_{m-1}=\frac{1}{m!},
\end{equation*}
 for all $j\in[m-1]$ (where $\mu$ denotes the standard Lebesgue measure).
\end{lemma}
\end{proof}

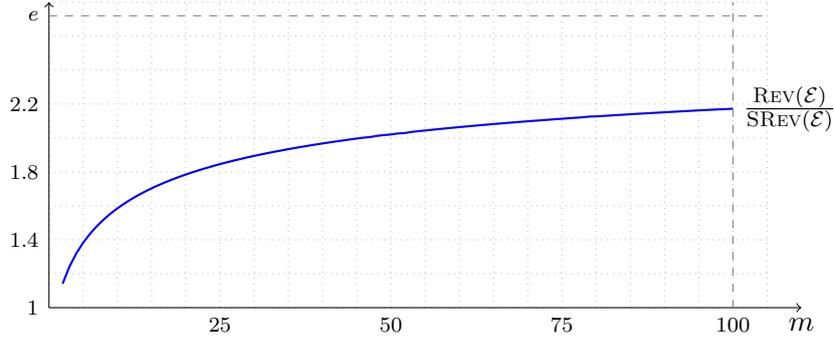
\begin{figure}
\begin{center}
\begin{tikzpicture}[domain=0:100,xscale=0.09,yscale=0.09]
\draw[dotted,step=5,very thin,color=gray] (0,0) grid (105,45);
  \draw[->] (0,0) -- (110,0) node[below] {$m$};
  \draw[->] (0,0) -- (0,45);
\draw[thick,color=blue] (2,3.54067) -- (3,6.05867) -- (4,7.99962) -- (5,9.57022) -- (6,10.8834) -- (7,12.0077) -- (8,12.9876) -- (9,13.8539) -- (10,14.6284) -- (11,15.3275) -- (12,15.9634) -- (13,16.5459) -- (14,17.0825) -- (15,17.5794) -- (16,18.0415) -- (17,18.4729) -- (18,18.8773) -- (19,19.2573) -- (20,19.6157) -- (21,19.9544) -- (22,20.2753) -- (23,20.58) -- (24,20.87) -- (25,21.1464) -- (26,21.4103) -- (27,21.6627) -- (28,21.9045) -- (29,22.141) -- (30,22.3604) -- (31,22.5733) -- (32,22.7797) -- (33,22.9768) -- (34,23.1608) -- (35,23.3568) -- (36,23.5345) -- (37,23.7077) -- (38,23.875) -- (39,24.0373) -- (40,24.195) -- (41,24.3477) -- (42,24.4962) -- (43,24.6405) -- (44,24.7802) -- (45,24.9169) -- (46,25.0493) -- (47,25.1445) -- (48,25.3209) -- (49,25.4277) -- (50,25.5193) -- (51,25.6641) -- (52,25.7199) -- (53,25.8898) -- (54,25.9987) -- (55,26.1052) -- (56,26.2091) -- (57,26.3105) -- (58,26.4104) -- (59,26.5086) -- (60,26.6036) -- (61,26.6974) -- (62,26.7886) -- (63,26.8788) -- (64,26.9665) -- (65,27.0533) -- (66,27.1381) -- (67,27.2206) -- (68,27.3021) -- (69,27.3826) -- (70,27.4618) -- (71,27.5377) -- (72,27.61) -- (73,27.6861) -- (74,27.7653) -- (75,27.8349) -- (76,27.9056) -- (77,27.9798) -- (78,28.0486) -- (79,28.1569) -- (80,28.1783) -- (81,28.2434) -- (82,28.3076) -- (83,28.3725) -- (84,28.4347) -- (85,28.4954) -- (86,28.5554) -- (87,28.6171) -- (88,28.6747) -- (89,28.73) -- (90,28.7897) -- (91,28.8443) -- (92,28.9035) -- (93,28.9555) -- (94,29.0095) -- (95,29.0647) -- (96,29.1186) -- (97,29.1668) -- (98,29.2211) -- (99,29.2472) -- (100,29.3211);
\node[right] at (100,29.3) {$\frac{\algoname{Rev}(\mathcal E)}{\algoname{SRev}(\mathcal E)}$};
 \draw[dashed,color=gray] (100,0) -- (100,45);
 \node[below] at (50,0) {\scriptsize$50$};
 \node[below] at (75,0) {\scriptsize$75$};
 \node[below] at (100,0) {\scriptsize$100$};
 \node[below] at (25,0) {\scriptsize$25$};
  \node[left] at (0,0) {\scriptsize$1$};
  \node[left] at (0,10) {\scriptsize$1.4$};
  \node[left] at (0,20) {\scriptsize$1.8$};
  \node[left] at (0,30) {\scriptsize$2.2$};
  \node[left] at (0,43) {\scriptsize$e$};
 \draw[dashed,color=gray] (0,43) -- (105,43);
\end{tikzpicture}
\end{center}
\caption{The approximation ratio bound in~\eqref{eq:expsepratio} for the separate selling mechanism for independent exponential valuations.}
\label{fig:exposepratio}
\end{figure}

An immediate result of Theorem~\ref{th:exppbound} combined with formula~\eqref{eq:expsep} is that for independent (not necessarily identical) exponential valuations the approximation ratio of the deterministic mechanism that sells items separately is at most
\begin{equation}\label{eq:expsepratio}
\frac{\revenue{\mathcal E}}{\algoname{SRev}(\mathcal E)}\leq\frac{G(m)}{m!}e<e.
\end{equation}
A plot of this approximation ratio $\frac{G(m)}{m!}e$ for the values $m=2,3,\dots,100$ can be found in Fig.~\ref{fig:exposepratio}. The loose constant factor bound $\frac{G(m)}{m!}<1$ is straightforward and the proof can be found in Appendix~\ref{append:funcg}.

\subsection{A Simple Randomized Selling Mechanism}
Consider the following very simple randomized mechanism for the setting of independent exponential valuations with parameters $\lambda_1,\dots,\lambda_m$. Without loss of generality, in the following let's assume that $\lambda_1\geq\lambda_2\geq\dots\geq\lambda_m$. We will again be using our notation of $w=\sum_{j=1}^m\lambda_jx_j$ and $\hat\lambda=\prod_{j=1}^m\lambda_j$.
\begin{definition}[Mechanism \algoname{Proportional}]\label{def:proportional}
Sell item $j$ with probability $\frac{\lambda_j}{\lambda_1}$ and collect a total payment of $\gamma^*_m/\lambda_1$ (parameter $\gamma^*_m$ is defined before~\eqref{eq:Gfunction}).
\end{definition}
Essentially we sell the items with probability proportional to their parameters, normalized by the largest parameter $\lambda_1$. This mechanism is truthful, because it corresponds to the following utility function 
$$
u(\vecc x)=\max\sset{0,x_1+\frac{\lambda_2}{\lambda_1}x_2+\dots+\frac{\lambda_m}{\lambda_1}x_m-\frac{\gamma^*_m}{\lambda_1}}
$$
which is obviously convex and we will use the shorthand notation
\begin{equation}\label{eq:proportionalu}
U(w)=u(\vecc x)=\max\sset{0,\frac{w}{\lambda_1}-\frac{\gamma^*_m}{\lambda_1}}
\end{equation}
when this is more comfortable. Now let's compute \algoname{Proportional}'s expected revenue. By~\eqref{eq:totalrevenue} and the fact that $f_j(x_j)=\lambda_je^{-\lambda_jx_j}$ this is
$$
\hat\lambda\int_{\R^m_+}\left(\sum_{j=1}^m\frac{\partial u(\vecc x)}{\partial x_j}-u(\vecc x)\right)e^{-\sum_{j=1}^m\lambda_jx_j}\,d\vecc x
$$
and by a simple integration by parts (see e.g. the derivation in~\citep[Sect.~2]{Daskalakis:2013vn}) this can be written as
$$
\hat\lambda\int_{\R^m_+}u(\vecc x)\left(\sum_{j=1}^m\lambda_jx_j-(m+1)\right)e^{-\sum_{j=1}^m\lambda_jx_j}\,d\vecc x=\hat\lambda\int_{\R^m_+}u(\vecc x)\left(w-(m+1)\right)e^{-w}\,d\vecc x
$$
and by performing the same change of variables as in~\eqref{eq:changevar1} in the proof of Theorem~\ref{th:exppbound} we get that \algoname{Proportional}'s expected revenue is 
$$
\hat\lambda\int_{0}^\infty\int_\mathcal{A}U(w)\left(w-(m+1)\right)e^{-w}\frac{w^{m-1}}{\lambda}\,dw\,dt_1\,\dots\,dt_{m-1}
$$
which equals
$$
\frac{1}{(m-1)!}\int_{0}^\infty U(w)w^{m-1}\left(w-(m+1)\right)e^{-w}\,dw
$$
by using Lemma~\ref{lemma:chainunit}. Utilizing~\eqref{eq:gderiv} and taking into consideration that $U(0)=0$ and that $\lim_{w\to\infty} g(m,w)U(w)=0$, integrating by parts the revenue becomes
$$
\frac{1}{(m-1)!}\int_{0}^\infty U'(w)g(m,w)\,dw=\frac{1}{(m-1)!}\int_{\gamma^*_m}^\infty \frac{1}{\lambda_1}g(m,w)\,dw=\frac{G(m)}{(m-1)!\lambda_1},
$$
for the first equation using~\eqref{eq:proportionalu}. So we showed the following:
\begin{mdframed}
\begin{theorem}\label{th:proportionalbound}
For $m$ goods with independent (but not necessarily identical) exponential valuations (with parameters $\lambda_1\geq\lambda_2\geq\dots\geq\lambda_m$), mechanism \algoname{Proportional} has an expected revenue of $$\frac{G(m)}{(m-1)!\lambda_1},$$
where $G(m)$ is defined in~\eqref{eq:Gfunction}.
\end{theorem}
Immediately, by combining Theorem~\ref{th:proportionalbound} with Theorem~\ref{th:exppbound}, we get that \algoname{Proportional}'s approximation ratio is upper-bounded by
\begin{equation}\label{eq:proportionalratio}
\frac{1}{m}\left(1+\frac{\lambda_1}{\lambda_2}+\dots+\frac{\lambda_1}{\lambda_m}\right)\leq\max_j\frac{\lambda_1}{\lambda_j}=\frac{\lambda_1}{\lambda_m}.
\end{equation}
\end{mdframed}
The performance of this approximation ratio bound depends heavily on the choice of the parameters $\lambda_j$. Obviously, the closer these parameters are the better the bound. However, if $\lambda_1\gg\lambda_m$ then this ratio can be unbounded. In such a case though, we can fall back to using the constant approximation separate selling mechanism in~\eqref{eq:expsepratio} which is at most $e$-approximate. A very interesting consequence of~\eqref{eq:proportionalratio} is for the special case of i.i.d.\ exponential priors, i.e.~when $\lambda_1=\dots=\lambda_m=\lambda$. In that case, by~\eqref{eq:proportionalu} it is straightforward to see that \algoname{Proportional} reduces to the simple deterministic mechanism that sells all items in a full bundle for a price of $\gamma^*_m/\lambda$ and also the approximation ratio in~\eqref{eq:proportionalratio} becomes $1$, meaning that full bundling is optimal:
\begin{mdframed}
\begin{theorem}\label{eq:expoiidbundleoptimal}
Selling deterministically in a full bundle\footnotemark \ is optimal for \emph{any number} of exponentially i.i.d.~goods.
\end{theorem}
\end{mdframed}
\footnotetext{The optimal bundle price is $\gamma^*_m/\lambda$, where $\lambda$ is the parameter of the exponential distribution and $\gamma^*_m$ is given before~\eqref{eq:Gfunction}.}

\paragraph{Acknowledgements:} I am deeply grateful to Elias Koutsoupias for many useful discussions and guidance.  I also thank an anonymous reviewer for suggesting the discussion after Theorem~\ref{th:optimaldualbounduniform}.

\bibliographystyle{abbrvnat} 
\bibliography{BoundingOptimal}

\appendix
\providecommand{\refgmwfunc}{\ensuremath{g(m,w)} }
\section{Properties of Function \protect\refgmwfunc of Section~\ref{sec:expoenential}}\label{append:funcg}
Fix some positive integer $m$. Then $g(w)\equiv g(m,w)$ is an absolutely continuous function on $\R_+$ with derivative
\begin{align}
g'(w) &=\frac{\partial \varGamma(m+1,w)}{\partial w}-(m+1)\frac{\partial\varGamma(m,w)}{\partial w}\notag\\
		&=-\frac{w^m}{e^w}-(m+1)\left(-\frac{w^{m-1}}{e^w}\right)\notag\\
		&= (m+1-w)w^{m-1}e^{-w}.\label{eq:gderiv}
\end{align}
This means that $g(w)$ is strictly increasing in $[0,m+1]$ and strictly decreasing in $[m+1,\infty)$. Also we can compute:
\begin{align*}
g(m,0) &=\varGamma(m+1,0)-(m+1)\varGamma(m,0)\\
		&=m!-(m+1)(m-1)!=-(m-1)!<0\\
g(m,m+1) &= \varGamma(m+1,m+1)-(m+1)\varGamma(m,m+1)\\
		&=(m-1)!e^{-(m+1)}\left[m\sum_{k=0}^{m}\frac{(m+1)^k}{k!}-(m+1)\sum_{k=0}^{m-1}\frac{(m+1)^k}{k!} \right]\\
		&=(m-1)!e^{-(m+1)}\sum_{k=0}^{m-1}\left(\frac{(m+1)^m}{m!}-\frac{(m+1)^k}{k!}\right)>0\\
\lim_{w\to\infty}g(m,w) &=0.
\end{align*}
From the above we can deduce that $g(w)$ has a unique root $\gamma^*$ in $\R_+$. In fact $\gamma^*\in(0,m+1)$ and $g(w)<0$ for all $w\in(0,\gamma^*)$ and $g(w)>0$ for all $w\in(\gamma^*,\infty)$.

Furthermore, we know that the incomplete gamma function has the property
\begin{equation}\label{eq:gammaprop1}
\varGamma(m+1,w) =m\varGamma(m,w)+w^me^{-w}.
\end{equation}
With the help of this we can see that
\begin{align*}
\int_a^\infty\varGamma(m,w) &=\left[w\varGamma(m,w)-\varGamma(m+1,w)\right]_a^\infty\\
		&=\left[w\varGamma(m,w)-m\varGamma(m,w)-w^me^{-w}\right]_a^\infty\\
		&=\left[(w-m)\varGamma(m,w)-w^me^{-w}\right]_a^\infty\\
		&=(m-a)\varGamma(m,a)+a^me^{-a}
\end{align*}
for any positive integer $m$ and $a\in\R_+$, so 
\begin{align*}
\int_a^\infty g(m,w)\,dw &= \int_a^\infty\varGamma(m+1,w)-(m+1)\int_a^\infty\varGamma(m,w)\\
		&=(m+1-a)\varGamma(m+1,a)+a^{m+1}e^{-a}-(m+1)(m-a)\varGamma(m,a)-(m+1)a^me^{-a}\\
		&=(m+1-a)(m\varGamma(m,a)+a^me^{-a})-(m+1)(m-a)\varGamma(m,a)+a^me^{-a}(a-m-1)\\
		&=\left[(m+1-a)m-(m+1)(m-a)\right]\varGamma(m,a)\\
		&=a\varGamma(m,a).
\end{align*}

Also due to~\eqref{eq:gammaprop1} we get
\begin{align*}
g(m,w) &=\varGamma(m+1,w)-(m+1)\varGamma(m,w)\\
		&=m\varGamma(m,w)+w^me^{-w}-(m+1)\varGamma(m,w)\\
		&=w^me^{-w}-\varGamma(m,w)
\end{align*}
and so, since $\gamma^*$ is a root of $g(m,w)$ this means that
$$
\varGamma(m,\gamma^*)=(\gamma^*)^m e^{-\gamma^*}.
$$

We will now show the bound $\frac{G(m)}{m!}<1$ we used in~\eqref{eq:expsepratio}. We have:
\begin{align*}
\frac{G(m)}{m!} &=\frac{\gamma^*_m\varGamma(m,\gamma^*_m)}{m!}&&\text{from~\eqref{eq:Gfunction}}\\
		&=e^{-\gamma^*_m}\frac{\gamma^*_m}{m}\sum_{k=0}^{m-1}\frac{{\gamma^*_m}^k}{k!}\\
		&< e^{-\gamma^*_m}\sum_{k=0}^{m-1}\frac{{\gamma^*_m}^{k+1}}{(k+1)!}&&\text{since}\;\;k+1\leq m\\
		&< e^{-\gamma^*_m}\sum_{k=0}^{\infty}\frac{{\gamma^*_m}^{k}}{k!}=1.
\end{align*}

\section{Remaining Computation in the Proof of Theorem~\ref{th:optimaldualbounduniform}}\label{append:uniform1}
\begin{align*}
\int_{I^m}\sum_{j=1}^mz_j(\vecc x)\,d\vecc{x} &=\sum_{\vecc v\in \mathcal I_m}\int_{L_{\vecc v}}\sum_{j=1}^mz_j(\vecc x)\,d\vecc{x}\\
		&=\sum_{\vecc v\in\mathcal  I_m}\int_{L_{\vecc v}}\sum_{j:\vecc v_j=1}z_j(\vecc x)\,d\vecc{x}\displaybreak[3]\\
		&=\sum_{\kappa=1}^m\sum_{\vecc v:k(\vecc v)=\kappa}\int_{L_{\vecc v}}\sum_{j:\vecc v_j=1}z_j(\vecc x)\,d\vecc{x}\displaybreak[3]\\
		&=\sum_{\kappa=1}^m\binom{m}{\kappa}\underbrace{\vphantom{\int_{\frac{1}{m+1}}^{1}\dots\int_{\frac{1}{m+1}}^{1}}\int_0^{\frac{1}{m+1}}\dots\int_0^{\frac{1}{m+1}}}_{\text{$m-\kappa$ times}}\underbrace{\int_{\frac{1}{m+1}}^{1}\dots\int_{\frac{1}{m+1}}^{1}}_{\text{$\kappa$ times}}  \sum_{j:\vecc v_j=1}z_j(\vecc x)\,d\vecc{x}\displaybreak[3]\\
		&=\sum_{\kappa=1}^m\binom{m}{\kappa}\int_0^{\frac{1}{m+1}}\dots\int_0^{\frac{1}{m+1}} \int_{\frac{1}{m+1}}^{1}\dots\int_{\frac{1}{m+1}}^{1}  \sum_{j:\vecc v_j=1}\frac{m+1}{\kappa}(x_j-c_k)\,d\vecc{x}\displaybreak[3]\\
		&=\sum_{k=1}^m\binom{m}{k}\int_0^{\frac{1}{m+1}}\dots\int_0^{\frac{1}{m+1}} \int_{\frac{1}{m+1}}^{1}\dots\int_{\frac{1}{m+1}}^{1}  \sum_{j:\vecc v_j=1}\frac{m+1}{k}(x_j-c_k)\,d\vecc{x}\displaybreak[3]\\
		&=\sum_{k=1}^m\binom{m}{k}\left(\frac{1}{m+1}\right)^{m-k}k \int_{\frac{1}{m+1}}^{1}\dots\int_{\frac{1}{m+1}}^{1} \int_{c_k}^1 \frac{m+1}{k}(x-c_k)\,dx\displaybreak[3]\\
		&=\sum_{k=1}^m\binom{m}{k}\left(\frac{1}{m+1}\right)^{m-k}k \left(1-\frac{1}{m+1}\right)^{k-1}\int_{c_k}^{1}  \frac{m+1}{k}(x-c_k)\,dx\displaybreak[3]\\
		&=\sum_{k=1}^m\binom{m}{k}\frac{m^{k-1}}{(m+1)^{m-2}}\int_{c_k}^1(x-c_k)\,dx\displaybreak[3]\\
		&=\sum_{k=1}^m\binom{m}{k}\frac{m^{k-1}}{(m+1)^{m-2}}\frac{(1-c_k)^2}{2}\displaybreak[3]\\
		&=\frac{1}{2(m+1)^m}\sum_{k=1}^m\binom{m}{k}k^2m^{k-1}\displaybreak[3]\\
		&=\frac{m(1+m^2)}{2(1+m)^2}.		
\end{align*}
\end{document}